\documentclass[useAMS,usenatbib,A4]{mn2e}
\newif\ifAMStwofonts
\AMStwofontstrue \voffset=-2cm
\def\lesssim{\ \raise -2.truept\hbox{\rlap{\hbox{$\sim$}}\raise5.truept  
\hbox{$<$}\ }}                              %
\def\gtrsim{\ \raise -2.truept\hbox{\rlap{\hbox{$\sim$}}\raise5.truept  %
\hbox{$>$}\ }}                              %

\title[The Impact of Energy Feedback on AGNs]{The Impact of Energy Feedback on \\
 Quasar Evolution  and Black Hole Demographics}
\author[V. Vittorini, F. Shankar, A. Cavaliere]{V. Vittorini$^{1}$, F.
Shankar$^{2}$, A. Cavaliere$^1$\thanks{E-mail:
victor@roma2.infn.it; shankar@sissa.it; cavaliere@roma2.infn.it}\\
$^1$ Astrofisica, Dip. Fisica, Universit\'a  Tor Vergata,
via Ricerca Scientifica 1, I-00133 Rome, Italy \\
$^2$ SISSA , Via Beirut 4, I-34014 Trieste, Italy}
\begin{document}
\date{         }
\pagerange{\pageref{firstpage}--\pageref{lastpage}} \pubyear{2005}
\maketitle
\label{firstpage}

\begin{abstract}
We investigate how accretion episodes onto massive black
holes power quasars and active galactic nuclei while they accumulate mass into the holes.
We implement an  analytic
approach to compute both the trend and the stochastic component
to the trigger of the accretion events, as provided by structure
buildup after the hierarchical  paradigm. We base on host galaxy evolution proceeding from the
protogalactic era at redshifts $z \gtrsim 2.5$ dominated by major
merging events in high density regions, to the subsequent era marked by galaxy-galaxy
interactions in newly forming groups. These dynamical events
perturb the gravitational equilibrium of the gas reservoir in the hosts, and
trigger recurrent accretion episodes first in the Eddington-limited
regime, later in a supply-limited mode controlled by energy
feedback from the very source emission onto the surrounding gas.
Depletion of the latter by these events (adding to quiescent star formation) concurs with
the slowing down of the clustering to cause a fast drop of the
activity in dense regions. Meanwhile,  in the ``field" later and rarer events
are triggered  by interactions of still gas-rich galaxies, and
eventually by captures of dwarf satellite galaxies; these are also included in our analytic model.
Thus we compute the quasar and AGN luminosity functions; we find these to brighten and rise
from $z \simeq 6$ to $z \simeq 2.5$, and then toward $z \simeq 0$ to dim and fall somewhat,
 in detailed agreement with the observations.
From the same accretion history we predict that for $z < 2.5$ the
mass distribution of the holes  progressively rises and shifts
rightwards;  we compare our results with the local data. We also
find that downward of $z  \simeq 2.5$ the  Eddington ratios related
to emitting, most massive holes drift below unity on average, with a
widening scatter; meanwhile, some smaller holes flare up closer to
the Eddington limit. We conclude that the accretion history, however
rich,  is dominated by  dwindling events triggered by interactions
under the control by feedback; these  establish a link between the
declining but widely scattered distribution of the Eddington ratios
and the tight, closely stable upper section of the $M-\sigma$
correlation. Two clearcut  predictions   arise  from our interaction
picture: the BH activity will appear to follow  an anti-hierachical
trend; the QSO-AGN population is expected to be bimodal, and related
to the bimodal galaxy population.

\end{abstract}

\begin{keywords}
black hole physics -- galaxies: active -- galaxies: interactions --
galaxies: nuclei-- quasars: general .
\end{keywords}

\section{Introduction}

Accreting black holes (BHs) with masses $M\sim 10^6\,-\,10^9\,
M_{\odot}$ are widely held (Rees 1984) to energize  Active Galactic
Nuclei (AGNs) and quasars (QSOs) and produce  their huge bolometric
outputs  that approach $L \sim 10^{48}$ erg s$^{-1}$ (e.g., Groote,
Heber \& Jordan 1989, Hagen et al. 1992, Maraschi \& Tavecchio
2003). Not only strong gravity but also large mass inflows
$\dot{M}=L\,/\,\eta c^2$ up to some  $10^2\, M_{\odot}$yr$^{-1}$
must be  involved in these objects, even when the overall efficiency
$\eta$ for converting gravitational into radiative energy is up to
10\%.

By the same token the BH masses
\medskip
\begin{equation}
M(t)=\frac{1}{\eta c^{2}}\int^t dt'L(t')
\end{equation}
\medskip
are expected to keep the archives of the luminous history of these
sources down to the cosmic time $t$; barring major silent accretion,
the above equation relates the \textit{cumulative} variable $M$ to
the quantity $L$ which signals \textit{current} source activity.

The integral nature of Eq. (1) allows three main  activity patterns, as originally
discussed by Cavaliere \& Padovani  (1988), see also Kembhavi \& Narlikar (1999). For one,  the QSO
population could be comprised of a limited number of sources continuously
active over several Gyrs, that accumulate large masses $M(t)$ in excess of
$10^{10}\, M_{\odot}$ while emitting lower and lower outputs $L(t)$,
to agree with the fall observed in the bright QSO population
for redshifts $z<2.5$ (see Osmer 2004 for a review). But then the ratios
$L/M$ ought to  drop  sharply with the cosmic time $t$; strong
$t$-dependence would also occur in the relation between the
increasing BH masses and the more persisting properties of their host
galaxies, e.g., the velocity dispersion $\sigma$. However, this
specific pattern is ruled out by the observations of supermassive
BHs widespread in many local and currently inactive galaxies, with
an upper mass envelope at $M \lesssim 5\,10^9\, M_{\odot}$
(see Tremaine et al. 2002); a similar bound is also found for QSOs shining at higher
$z$ (see McLure \& Dunlop 2003, Vestergaard 2004).

At the other extreme, Eq. (1) holds as well during a single
accretion event of a mass $\mu$ over the time scale $\tau \sim
10^{-1}$ Gyr to yield a flash of bolometric luminosity $L\simeq \eta
c^2\mu/\tau$. Here each BH's mass is  accreted in one go over a
constant $\tau$, so $M = \mu$ applies and the ratios $L/M \simeq
\eta c^2\,/\,\tau$ ought to be  closely constant. Here  the QSO fall
would be made up by many sources active only once and briefly,
conceivably self-limited by the radiation pressure at the Eddington
value $L = L_E\equiv Mc^2/t_E$, with $M(t)$  exponentiating  on the
Salpeter timescale $\eta t_E \simeq 5\,10^{-2}$ Gyr. Over the cosmic
time $t$ new BHs ought to form continuously, but with progressively
lower masses for $z < 2.5$ so as to track the observed population
decline to lower luminosities while retaining constant Eddington
ratios $\lambda_E = L/L_E  \simeq 1$. For this to occur, the trend
in the hierarchical formation of structures toward ever-growing
masses would have to be reversed in a closely tuned way for the
active BHs (see discussion by Merloni 2004). Then the tight
$M-\sigma _c$ correlation observed between the BH masses $M$ and the
bulge velocity dispersions $\sigma_c$ (first pointed out by
Ferrarese \& Merritt 2000, Gebhardt et al. 2000) would evolve
markedly, being progressively extended toward smaller $M$ for
decreasing $z$, still with a narrow scatter.

Finally, the luminous history  may involve in many  bright galaxies  a
number of \textit{recurrent} accretion events with decreasing
outputs; then Eq. (1) yields
\begin{equation}
L_k\simeq \frac{\eta c^2\mu_k}{\tau}~,
\end{equation}
and the BH  mass has the form $M=\sum_{k}\mu_{k}$ with the few first
contributions dominating. Now the individual  BH's masses will grow
moderately with $t$,  in accord with the intrinsic hierarchical trend but
in accord also with the decreasing average luminosities. Such
\textit{limited supplies}  will cause  a mild drift of the average Eddington
ratios  below the value $\lambda_E \simeq 1$, but only weak
evolution in the $M-\sigma_c$ relation. Here we plan to discuss the
latter two activity patterns, motivated by a number of observations.

Shields et al. (2003) observe the $M-\sigma_c$ correlation
to be largely in place at epochs $z\simeq 2.5$ at least in its upper range,
the  masses being afterward modified  by factors
of a few at most. The single-flash pattern of activity with $\lambda_E
\simeq \,$const hardly accords with these findings, see discussion by
Corbett et al. (2003); it also disagrees with the considerable
scatter shown by ratios $\lambda_E$.

As to the latter, Eddington luminosities are welcome at the
highest observed redshifts $z > 6$, when the age of Universe was
definitely shorter than 1 Gyr and yet powerful QSOs with  $L\sim
10^{47}$erg s$^{-1}$ already appear in the SDSS data (Fan et al.
2003). These sources require minimal BH masses $M \simeq 10^{9}
M_{\odot}$ if they shine at $\lambda_E \simeq 1$, a condition that
also fosters fast growth from much smaller seeds. In fact, for $z
\simeq 2.1 $ McLure \& Dunlop (2003) find sources with Eddington ratios
often attaining values $\lambda_E \simeq 1$ (see also Corbett et al.
2003), but undergoing a slow average decrease toward $z \simeq
0.2$. In addition, Vestergaard (2004) finds $\lambda_E$ to show
large, partly intrinsic scatter at all redshifts $z \lesssim 4$,
with a declining upper envelope for the massive objects as $z$
decreases. These observations indicate that the radiation pressure
limit  to accretion is often attained at high $z$, but also that it
does not always constitute the tightest constraint at lower $z$. We
propose that for $z \lesssim 2$ a {\it supply} limit may be more
effective for many BHs.

The counterpart of Eq. (1) extended to the entire unobscured population (or populations) is
provided by the relation (cf. Soltan 1982)
\medskip
\begin{equation}
\int dM  N(M,t)\,M=\frac{1}{\eta c^2} \int^{t} dt' \int dL\, N(L,t')
\,L ~.
\end{equation}
\medskip
Here $N(L,t)$ is the bolometric luminosity function (LF) of the
quasars and AGNs, and $N(M,t)$ is the mass distribution (MD) of the
related  BHs.

The development of the former is known to be
sharp, coherent and non-monotonic;  the QSO luminosity density peaks at the
epoch corresponding to $z \simeq 2.5- 3$, and  rises and falls on
both sides on timescales of a few Gyrs (Schmidt 1989, Osmer 2004).
In detail, the optical LF ``evolves" sharply
from the Local Universe, with the number of luminous  QSOs rising
out to $z \simeq 2.5$ (see Grazian et al. 2000; Croom et al. 2004).
Past this redshift,  ``negative evolution" occurs
with the luminous QSOs dwindling on average  out to $z  \simeq 5$ and
beyond (see Kennefick, Djorgowski \& De Carvalho 1995; Schmidt,
Schneider \& Gunn 1995; Fan et al. 2001). A similar trend is also
shown by the LF observed in the X-ray band out to $z \simeq 3$ (see
Miyaji, Hasinger \& Schmidt 2000; La Franca et al. 2001; Cristiani et al. 2004a).

Less known is the  behaviour of the MD. The local distribution has
been estimated from the observations of a number of relic
supermassive BHs (see Yu \& Tremaine 2002; Marconi et al 2004;
Shankar et al. 2004). These authors find the integrated mass density
$\rho_{bh}\, \equiv \int dM\,  N(M,t_o)\, M$ to take on local values
in the range $ 4-6\; 10^5 M_{\odot}/$Mpc$^3$,  consistent after Eq.
(3) with the optical emissions from QSOs summed to the X-ray
emissions from AGNs, with a considerable
contribution from obscured sources. In the range $z < 2.5$ where the quasar
population declines, these AGNs still contribute appreciably to the BH
masses; eventually,  the dominant contribution comes from the
AGNs pinpointed in  X-rays  (Hasinger 2003, Fabian 2004, Cristiani et al. 2004a).

We stress that Eqs. (3) and (1) with their integral nature require
neither the variables $L$ and $M$ to be linked by a constant value
of $\lambda_E$, nor the functions  $N(M,t)$ and $N(L, t)$  to  be
linked by a relation of the type $N(L, t)\,dL \propto \delta\,  N(M,t)\, dM
 $ in terms of the emission duty cycle $\delta \sim \tau/t$.  In fact, both
assumptions, however appealingly simple, run at some point against a
piece of the available data, especially on our side of the QSO peak.
In this paper we shall \textit{relax} these assumptions, and will
widen our scope  to derive the actual relationship between $L$ and
$M$ and between $N(L, t)$ and $N(M,t)$; in \S 2 we give our
guidelines. In \S 3 we describe how supermassive BHs form and shine
at $z>2.5$ while their protogalactic hosts are built up. In \S 4 we
compute how formed BHs are re-fueled  during the subsequent era $z
<2.5$ by interactions of their hosts with companion galaxies in
dense regions. In \S 5 we develop our formalism to evolve the BH
mass function throughout the two above eras. In \S 6 we derive the
QSO luminosity function evolving under the drive of the above BH
accretion history. In \S 7 we consider the late fueling events due
to interactions in low density regions and to captures by the host
of satellite galaxies. In \S 8 we sum up and discuss our findings.

Our framework will be provided by  the ``Concordance Cosmology" with
$H_{0}\simeq 70$ km/s~Mpc, $\Omega_{m}\simeq 0.27,\;
\Omega_{\Lambda}\simeq 0.73,\; \Omega_{b}\simeq 0.044$ (see Bennett
et al. 2003); this implies the relation we use between $t$ and $z$,
namely, $t =11 \; sinh^{-1} \,[1.6 \, (1+z)^{-3/2}]$ Gyr.

\section{Our guidelines}

Many authors have attempted to model the connected  evolution of
the  BH MD and of the shining QSO LF, particularly:
Cavaliere \& Vittorini (2000 and 2002, CV00 and CV02 hereafter),
Kauffmann \& Haehnelt (2000), Cattaneo (2001); more recently yet,
Volonteri, Haardt \& Madau (2003), Menci et al. (2003), Wyithe \& Loeb
(2003), Steed \& Weinberg (2003), Marconi et al. (2004), Granato et al. (2004), Merloni (2004).

In the present paper our guideline will be that  such a connection has to take  place through
the Eqs. (1) and (3) intrinsic to any accretion process. These
processes are far from trivial, since to power  sources as luminous
as the bright QSOs they have to involve large gaseous masses and
high accretion rates. For outputs exceeding $L \sim 10^{47}
\,L_{\odot}$ erg s$^{-1}$, masses $\mu \sim 10^{9}M_{\odot}$ or more
must be funnelled toward the central BH in  times $\tau \sim
10^{-1}$Gyr or less. In addition, these accretion episodes must be
coordinated into the highly coherent, non-monotonic evolution the
QSOs exhibit on scales of a few Gyrs over a span of about 13 Gyr.

Large gravitational perturbations involving a considerable volume of
the host galaxies, such as  merging of protogalaxies or galaxy-galaxy
interactions, are required in order to distort the galactic
potential on kpc scales, and remove enough angular momentum from the
galactic gas so as to start it on an inward course toward the central
BH.

But major mergers of the dark matter (DM) host  haloes as the only accretion triggers
hardly can account for the observed strong evolution of QSOs for
$z\lesssim 2.5$. In fact, as discussed in detail by Menci et al.
(2003), the widely agreed paradigm of hierarchical growth of cosmic structures implies
the merging rate at galactic scales to decrease slowly  by a factor of some
$10^{-1}$ from the QSO peak to the local Universe.  This by itself
would lead to overpredict the bright QSOs surviving at low  $z$, at
variance with the dramatic fall observed in the QSO population which  entails  factors  of
order $10^{-2}$ in the number of bright sources.  Even when
account is taken of the galactic gas exhaustion due both to
BH accretion itself and to thhe ongoing  star formation (see Kauffmann \&
Haehnelt 2000, Cattaneo 2001), the computed population of QSOs with
blue magnitudes $M_B\le -24$ would still outnumber that observed for
$z\lesssim 1$.  Moreover, the  MD accordingly computed differs
considerably at $z \simeq 0$ from what is inferred on the basis of the $M-\sigma$
observations and the distribution of the velocity dispersions (see
Wyithe \& Loeb 2003).

We instead relate BH growth with QSO and AGN luminosities from
considering the rich picture that arises naturally after  the
hierarchical formation of cold DM structures; this envisages the initially minute density perturbations
to grow, collapse and virialize under the drive of the gravitational instability, and in closer detail to
merge with similar clumps into larger structures (see Peebles 1993). We follow this standard course, and
consider how the various dynamical events it predicts
affect the dynamical life of the host galaxies and trigger different modes of accretion
onto the central BH.

These events include the early major {\it mergers} that assemble the massive protogalaxies, but
also include later and milder tidal {\it interactions} of the formed hosts
with companion galaxies in dense environments like groups, or even in lower density environments such
as  the Large Scale Structures (LSS);
eventually, the hosts cannibalize their retinues of satellite galaxies.
The interactions in dense environments (groups in particular),  while still able  to
trigger large inflows of the galactic gas toward the nucleus, outnumber the
 bound mergers for $z < 2.5$, consume more gas and
decay faster, so speeding up the QSO evolution (see Menci et al.
2003).

Later on for $z < 1 $ ``field" processes such as interactions of gas-rich hosts in LSS and
satellite cannibalism together are left as the  dominant, if often meager fueling  mode,
so as to be  phenomenologically perceived as a
later AGN population as opposed to that related to dense environments.

We let all these dynamical events, with their overall \textit{trend}
and their \textit{stochastic} component, to form or rekindle  the
BHs as they may; we just record the outcomes from the integral
relationship  Eq. (1) of $M$ with $L$. As discussed in detail in \S
4, our key parameter here will be the gas fraction $f$ destabilized
by an  interaction,  with values around  $10\%$ based on transfer of
angular momentum from the orbital motion of the partner to the host
gas, and confirmed by numerical simulations.

We also compute
$N(M, t)$ and $N(L,t)$ separately but consistent with the  integral relation Eq. (3),
and compare these distributions with the observed ones.
Such  observable will be related to the statistics of the interactions based upon the mass
distributions of galaxies in groups, for which we adopt the standard Press \& Schechter (1974)
mass function.

We stress recent, direct evidences of activity connected with clearly
interacting galaxies given by Rifatto et al. (2001), Komossa et al.
(2003),  Ballo et al. (2004), and Guainazzi et al. (2005),  who report   AGNs  hosted in
both galaxies of the interacting systems ESO 202-G23, NGC 6240,
Arp 299, and ESO509-IG066, respectively. These findings complement the extensive, long known body
of evidence indicating that some 30\% QSO and strong AGN hosts have close companions or show
 signs of ongoing interactions, see the many single observations referred to in CV00, and
the statistics by  Bahcall et al. (1997) and by Kauffmann et al. (2003).
The evidence is even more significant in view of the different times conceivably
taken by inner fueling and by outer  disturbances, see the discussions by Beckman (2001) and by
Tadhunter et al. (2005), with the references therein.

One feature specific to the present work is our structural inclusion of the effects from the
source output itself onto the BH; we investigate how such an energy \textit{feedback}
from QSO emissions onto the surrounding gas can regulate the amount actually accreted and stored
into the  BHs. In fact, accretion is expected to be impaired or utterly halted
when the gas binding energy in the host is balanced by the energy
deposited by the source into the surrounding gas during a dynamical
time, see  Silk \& Rees (1998).

Whenever the accretion is so \textit{feedback-constrained}, the
balance condition clearly translates into a steep $M -\sigma$
relation between the mass accreted and the depth of the host
potential well. In closer detail,  we will compute how the feedback
affects the evolution of the $M-\sigma$ relation, the Eddington
ratios, and  the evolving LF and MD. As discussed in detail in \S 3,
the key parameter here will be the feedback efficiency $\phi$ to
deposit source energy into the host gas; we will use values $\phi
\sim  v/2c \simeq$ a few $\%$ following from momentum conservation
from radiation to gas,  and consistent with independent lines of
data.

\section{Forming BHs in protogalactic haloes for $z\gtrsim 2.5$}

Taking up from CV02, it is convenient to divide  the growth of the
BHs into two main regimes that overlap around $z\simeq 2.5$, with
later additions.

At early epochs $z >  2.5$, supermassive BHs grow mainly during the
major \textit{merging} events that in dense environments build up massive protogalactic haloes of
masses $M_{h} \sim  5\,  10^{11}\,-\, 10^{13} M_{\odot}$. By these
events large amounts of gas are destabilized and funnelled towards
the galactic centre to be eventually accreted onto the BH.
Meanwhile, the same events also replenish the host structures with
fresh gas supplies, and so sustain the amounts $m$ of galactic gas at nearly
cosmic levels $m/M_h  \simeq  \Omega_b/\Omega_m \simeq 0.15$. As a result,
the accretion is often {\it self-limited} at nearly Eddington rates
with $L\propto M$, while protogalactic haloes and  BHs grow together
but not necessarily in a proportional fashion.
Thus  during  this  era it is appropriate to assume that BHs form with
masses $M = M_{in}$ directly related to the mass $M_{h}$ of the galactic
hosting halo.

We may express $M_h$ in terms of the DM velocity
dispersion $\sigma = A\,(GM_{h}\,/\,R)^{1/2}$, noting that for the
standard isothermal sphere $A \simeq  0.7$ applies, while for the DM
profiles by  Navarro, Frenk \& White (1997) a similar value $A \simeq
0.6 $ holds at the  virial radius R. To actually relate $M_{in}$ to
$\sigma$, we will consider two different  models.

$\bullet$ The \textit{unconstrained} accretion  model (UA) focuses on BH
coalescence directly following the merging of their host haloes; this
yields the simple proportionality $M_{in}\simeq 10^{-4} M_{h}$ (see
Haehnelt \& Rees 1993, Volonteri, Haardt \& Madau 2003). When expressed in
terms of $\sigma$ in units $\sigma_*=200\,$km s$^{-1}$ this yields
\begin{equation}
M_{in} \simeq 2\, 10^{-4}\,G^{-3/2}\, \sigma ^3\,
\rho^{-1/2}(z) \simeq
3\,10^7M_{\odot}\left(\frac{\sigma}{\sigma _*}\right)^4\,.
\end{equation}
The first relation $M_{in}\propto\sigma^3$ obtains at a fixed
virialization epoch, considering that $\sigma \propto M_h^{1/3}
\rho^{1/6}$ holds in terms of the DM density $\rho$ in the haloes.
The steeper course on the r.h.s. obtains from considering (Haehnelt \&
Kauffmann 2000) that the $z$-dependence may be approximately
rephrased in terms of an additional $\sigma$ dependence (plus a wide
residual scatter), because at high $z$ the standard hierarchical
scaling $\sigma \propto \rho^{(n-1)/6 (n+3)} \simeq \rho^{-1/2}$
applies, having used the index $n \simeq -2$ for the power spectrum of initial
density perturbations on galactic scales. When translated in terms
of of the stellar velocity dispersions  $\sigma_c
\propto \sigma^{1.2- 1.1}$ in the galactic bulges as indicated by Ferrarese (2002), by
Baes et al. (2003) and by Pizzella et al. (2004) the resulting average relation turns out to be
too flat compared with the
slope observed by Ferrarese \& Merritt (2000), Gebhardt et al. (2000),
while the scatter is still too large. Note that a similar argument
extended to $ z <1$, when the Concordance Cosmology modifies the
scaling laws into $\sigma\propto \rho ^{(1+n)/6(n+3)}$, would yield
instead too steep a relation $M_{in}\propto\sigma_c^{5}$.

$\bullet$ These drawbacks leads us to consider  the alternative model  based on
\textit{feedback-constrained} accretion (FCA) that includes the
energy balance  suggested by  Silk \& Rees (1998) and worked out by Haehnelt,
Natarajan \& Rees (1998); analogous results come from more detailed
work, e.g., King (2003), Granato et al. (2004), and  Lapi, Cavaliere \& Menci  (2005). The
balance condition reads
\medskip
\begin{equation}
\phi\,  L_{E}\,  t_{d} \simeq \frac{GM_{h}}{2 \, R}\; m\,~.
\end{equation}
\medskip
This applies since  gas unbinding and outflow occurs and accretion is halted when the fractional
quasar output $\phi$ deposited within a dynamical time $t_{d}\sim R/\sigma\simeq 10^8$ yr into
the current gas mass $m(t)$ in the host exceeds the binding energy of the
gas.  We begin with assuming  effective  values  $\phi\simeq 10^{-2}$ based
for radio-quiet QSOs upon  momentum conservation   between radiation and gas that yields
(in the absence of cooling) values up to   $ v/2c \simeq$ a few $\%$; for radio-loud QSOs
the kinetic energy in the jets
affords higher efficiencies, but the statistics of such sources is down to 10$\%$,
so conserves the weighted  value. The latter is independently confirmed on considering
(Lapi, Cavaliere \& Menci 2005) its  effects on the density of, and the X-ray
emission from hot gas in groups of galaxies surrounding the quasar hosts.
We also assume $\phi$ to be independent of the host mass, a point to be discussed in \S 8.

Recall now that  the hosts are resupplied with fresh gas under halo merging; so as long as the
overall gas  mass retains nearly cosmic values $m \simeq M_h\,
\Omega_b/\Omega_m \, \propto\sigma^3 $, the balance condition straightforwardly  yields
\medskip
\begin{equation}
M_{in}\simeq \frac{\Omega_b}{\Omega_m}\frac{t_E}{2\,  G\,  A^5\, c^2 \, \phi }
\,\sigma^5 \simeq
3\,10^7\, M_{\odot}\left(\frac{\sigma}{\sigma_*}\right)^5 ~.
\end{equation}
\medskip
In terms of $\sigma_c$, this is close to the observed slope
(the prefactor at $z\simeq 0$ after additional  mass has been accreted is
discussed below and given in Eq. 30).
Eq. (6)  may be recast into the form
$M_{in} =  6\,  10^{-6}\, (1+z)^{5/2}\, (M_{h}/10^{13}M_{\odot})^{2/3}\, M_h$,
to  make clear that feedback constrains  smaller haloes to form or grow BHs with
a  \textit{lower} efficiency $M_{in}/M_h \propto (M_h/10^{13}\, M_{\odot})^{2/3}$.

In both models,  during this era the QSO  outputs --
sustained at Eddington levels -- are  distributed after
a LF  directly related to the halo MD  by
the relation $N(L,t)\,dL= \tau\,\partial_t^+
N(M_{h},t)\,dM_{h}$; this is in terms of the hierarchical halo formation rate
$\partial_t^+ N (M_h, t) \propto  [M_c(t)/M_h]^{(n+3)/3} \; N(M_{h},t)/t $,
which decreases toward masses smaller than the average mass $M_c (t)$ virializing
at $t$ (see Cavaliere, Colafrancesco \& Scaramella 1991), so that  the r.h.s.
only approximately reads $ N_h(M_{h},t) \,dM_{h} \;\tau/t$.
By the same token, the hole MD ~ $N(M_{in},t)$ is also linked as given  by
\begin{equation}
N_{in}(M_{in},t) \, dM_{in}=N_{h}(M_{h},t)\,d M_h =
N_{h}(\sigma, t)\,d\sigma ~
\end{equation}
to  $N_{h}(M_{h}, t)$, or to  the equivalent $\sigma$
distribution  $N_{h}(\sigma, t)$. As for the former, we use the simple expression first
proposed by Press \& Schechter (1974) updated to the  Concordance Cosmology.
The MD computed from  Eq. (7) constitutes  at $z\simeq 2.5$
an initial condition to be evolved afterward under the drive of the interactions
described in \S 4.

The LF at early $z$ had been preliminarily  computed and discussed
by CV02. They stressed that for  $z >3 $ the FCA model  with the
ensuing non-linear stretching $L\propto M_{in} \propto M_{h}^{5/3}\,
(1+z)^{5/2}$ yields LF shapes  generally flatter than their UA
counterparts, and  more in tune with the then existing data at high
$z$ and bright $L$ (Fan et al. 2001). At fainter $L$ the present
model -- with the smaller prefactor corresponding to Eq. (6),
consistent with the additional accretion we envisage at later $z$ --
yields an even \textit{lower} LF, in tune with the recent
observations by Cristiani et al. (2004b), see Fig. 6. If needed, the
LF may be fine-tuned on  using the Sheth \& Tormen (1999) rendition
of the halo MD instead of the simple Press \& Schechter (1974) form,
and on adjusting the still unsettled amplitude of the initial
perturbation spectrum. In sum, three main \textit{intrinsic}
features concur to limit the number of  small BHs active at early
$z$: first, the lower efficiency in forming smaller haloes given by
$\partial_t^+ N(M_{h},t)$, yielded by  the hierarchical formation;
second, the lower efficiency of these in forming BHs after
$M_{in}/M_h \propto M^{2/3}_h$, due to  the feedback process; third,
the low prefactor that leaves room for the later increase of the BH
masses computed next.

\section{Refueling BHs in interacting hosts for $z < 2.5$}

For $z < 2.5$ major mergers become rarer and rarer at galactic
scales; fewer new massive BHs are formed at these epochs so their
number is conserved to a first approximation (improved in \S 7),
but their mass can still grow. Now the
prevailing dynamical events that trigger accretion are best
described as \textit{interactions} between developed galaxies, and
these occur mainly in the small, dense groups that at these epochs begin to
virialize. By the same token, the gas mass $m(z)$ in the hosts is
consumed with no  fresh imports provided by mergers. So the
accretion becomes {\it supply-limited} and can be easily
\textit{sub}-Eddington.

Small groups with mass exceeding $10^{13}\, M_{\odot}$, radius $R_G$
and bright galaxies membership $N_g
\gtrsim 3$ provide particularly suitable sites for the hosts to
interact with their companions (CV00); in fact, in early groups the
density $n_g$ of galaxies is high, while their velocity dispersion
$V$ is still comparable with the galactic $\sigma$, conditions that
favour effective binary interactions. These are mainly in the form
of {\itshape fly-by}, that is, binary encounters with impact
parameter $b$ ranging between the galactic radius $R$ and the value $R_G\,
N_g^{-1/3}$, that need not lead  to bound mergers. The cross
section is still close the geometrical value $\Sigma \simeq 4\pi
R^2 $ as long as $V \lesssim 2\, \sigma$ applies , and the average
time between these events is given by
\begin{equation}
\tau_r = 1/n_{g} \,\Sigma\, V ~;
\end{equation}
the local value is $\tau_{ro} \simeq 2$ Gyr. As groups merge into
rich clusters, $n_g$ decreases strongly following the  density $\rho(z)$ in the
host haloes, a trend only partially offset by the limited
increase of $V$. As a result,  the interaction rate $\tau_r^{-1}(t)$
declines with time following  $\rho\, V\propto (1+z)^{1.6}$ in the
Concordance Cosmology.

An  interaction of the host with a group companion of mass $M'$ will
perturb the galactic gravitational potential, and destabilize a fraction $f$ of the
cold gas mass $m$ in the host from its equilibrium at $r\sim$ kpc
from the centre. The amount $f\,m$ funneled to the galaxy centre ends
up in part into circumnuclear starbursts, and in a smaller part
trickles down to the accretion disk ending up onto the central BH. When the main constraint  governing
the gas equilibrium is provided by  the angular momentum $j$, the fraction $f$
may be computed as in CV00  to read
\medskip
\begin{equation}
f\simeq  |\frac{\Delta j}{j}|
\simeq A\, G\; \frac{M'}{b\, V\,\sigma}.
\end{equation}
\medskip
This  ranges  from some $f_{min}\simeq 5\,10^{-2}\,
(\sigma_*/\sigma)$  to $f_{max}\simeq 1/2$; the latter
constitutes the expected maximal gas fraction driven into the
central $10^2$ pc, as confirmed by aimed numerical simulations of galaxy interactions
(see  Mihos 1999). Corresponding to larger galaxies being more resistant to
gravitational distortion, Eq. (9) shows $f$ to scale as
$\sigma^{-1}$.

A fraction around $\sim 1/10$ of the inflowing mass
reaches the BH rather than ending into circumnuclear starbursts, as
indicated by the statistics of the energy sources that heat up the
dust in bright IR galaxies (see Franceschini, Braito \& Fadda 2003).
So $\mu = fm\,/\,10$ is the mass made available
for actual accretion, while the rest ends up into stars or is
dispersed. The process of fueling takes times of order $t_d$, the
host dynamical time, and spans  a few Salpeter times.

Between the limits $f_{min}, f_{max}$ the probability
density for $f$ due to the distribution of the orbital parameters
primarily reflects the distribution of the masses $M'$ of the interaction partners;
this is because the encounter velocity $V$ and the impact parameter
$b$ vary in narrow  ranges,   and are actually correlated in a galaxy group. So
$f \propto M'$ closely applies; since  $M'$ is distributed after
$p(M') \propto M'^{-s} $ with $s$ slightly under $2$, following
the Press \& Schechter (1974) distribution in its  power-law section, the
result is close to $p(f)\propto f^{-2}$. Th result  reads
\medskip
\begin{equation}
p(f | \sigma)=\frac{f_{max}\,f_{min}}{f_{max}-f_{min}}\,\, f^{-2}
~~~for~~~f_{min} \leq f \leq f_{max}~     .
\end{equation}
\medskip
This is used to compute the average value $\langle f \rangle \simeq
15\%$; since $f\propto \mu$ applies at given $m$, the result closely
reads   $p(\mu)\propto \mu^{-2}$.

Recurrent interactions will iteratively
\textit{exhaust}  the initial gas mass $m_{in}$ in the host; after $q$
interactions the residual  mass reads
\begin{equation}
m_q = m_{in}\,\prod^q_{k=1} (1-f_k)~,
\end{equation}
where each $f_k$ is extracted from the probability distribution Eq.
(10).

We will often use the simple estimate for the average  value
$\langle m_q \rangle \simeq m_{in}\, (1-\langle f\rangle)^q$, based
on equal average depletion factors $\langle f \rangle$; this is accurate to within $0.05 \,
m_{in}$ up to $q=7$ steps as we shall find below to occur on average from $z
\simeq 2.5$ to $0.2$.
We may also write for the average depletion rate the equivalent  differential
equation
\begin{equation}
\dot{m}/m \simeq -\, \langle f\rangle /\tau_r ~.
\end{equation}
In fact, its discretized solution starting from the initial condition $ m_{in}$
reads  just as $\langle m_q\rangle $ above when computed at the step
$q$ given by  the integer part of
\begin{equation}
 q = [\int^{t}_{t_{in}}
dt / \tau_r (t)]~,
\end{equation}
starting from $t_{in}$ that corresponds  to $z =2.5$; the r.m.s.
deviation  of the stochastic variable $q$ is about $\sqrt{q}$ to a
sufficient approximation. To the  depletion given by Eqs. (11) or
(12) we systematically add the gas consumption by ongoing star
formation, that we compute following Guiderdoni et al. (1998).

As for  the mass actually accreted, in this regime of
\textit{supply-limited} accretion we again consider two
possibilities.

$\bullet$ The \textit{unconstrained} accretion model (UA), where the
accretion is not affected by dynamical feedback, but still
are  subject to the radiation pressure limit. The mass accreted
in each interaction is $\mu = fm(z)\,/\,10$, distributed as given by
the simple counterpart of Eq. (10), namely
\begin{equation}
p(\mu  | \sigma)=g\;\mu^{-2}~~~for~~~f_{min}m/10\leq \mu\leq
f_{max}m/10 ,
\end{equation}
where $g = 10^{-1}\, m\,f_{max}\, f_{min}\,/(f_{max}-f_{min})$ provides the
normalization.

$\bullet$ The \textit{feedback-constrained} accretion model (FCA), in which
a tighter constraint to $\mu$ is set by the analogous of Eq. (5).
This is evaluated on considering that now the accretion may be
sub-Eddington, so $L_{E}$ is to be replaced by $ \eta
\,\mu\,c^{2}\,/\,\tau$, to read $\mu/m \simeq $  $ (\sigma/c)^2\,
\tau\,/\,2\, A^2\, t_d  \, \eta \,\phi$; so the constraint reads now
$ \mu \lesssim  \mu_{l} \equiv  m \;  \, (\sigma /c)^2 \, (\eta \,
\phi )^{-1}  $, considering that $\tau \simeq \,  t_d$ and
$A^2\simeq 0.5$ apply.

The mass actually accreted is given by the minimum between the
amount  $\mu = f \, m\,/10$ made available by the interaction, and
the constraint set by the feedback; that is to say,
\begin{equation}
\mu=min \left[  m\,f /10\, , \,\mu_{l}\right].
\end{equation}
After $q$ interactions $m$ is rescaled down iteratively as said, to
yield the estimate $\mu_{l}= M_{in} (1-\langle f\rangle)^{q}\,\tau
/\eta\,  t_E $. Note that the total probability of  the value
$\mu_{l}$ is contributed by all interactions leading to accretion
events that, if unconstrained, would exceed this value; this leads
to piling up of accretion episodes at  the upper bound  $\mu_{l}$.
To account for this,  we write the counterpart of Eq. (14) in the
form
\begin{eqnarray}
\nonumber
p(\mu\,|\,\sigma)&=&\delta(\mu-\,\mu_{l})~~~~~~~~~~~~~~~~~~~\,for~~~10\,
\mu_{l}\le f_{min}\, m \\
&=&g[\mu^{-2}+\,g'\; \delta(\mu-\mu_{l})]~~~~~otherwise
\end{eqnarray}
\medskip
where $g'\equiv (f_{max}\,  m-10\, \mu_{l})\,/\,(  f_{max}\, m
\;\mu_{l}) $.

In all cases the luminosity $L\propto\mu$ attained in any one
accretion  event  no longer is in a fixed relation to the current BH
mass $M$. This is because the accreted mass $\mu$ depends now on
\textit{stochastic} orbital parameters  as given in Eq. (9), and  is
distributed according to Eq. (14) or Eq. (16) in the UA or the FCA
model, respectively.

We end this Section  by giving a simple estimate of the final mass
of a BH after a number $q$ of interactions; when the feedback
constraint given by Eq. (15) is effective so as to join smoothly at $z=2.5$
with the mass $M_{in}$ similarly constrained by Eq. (6),
a simple upper bound is given by
\medskip
\begin{eqnarray}
\nonumber M\leq  \sum_{k=0}^{q}\mu_k &=& M_{in} + M_{in}\,
\sum_{k=1}^{q}(1-\langle f\rangle)^k=\\
&=& M_{in} \, [1-(1-\langle f\rangle)^{q+1}]/\langle f\rangle ~.
\end{eqnarray}
\medskip
In fact, this is close to the actual value in small and intermediate
haloes when the feedback  constrains $\mu$ to
be close to $\mu_{l}$, see  Eq. (15).

With  $\langle f\rangle = 15\%$  we find $M \lesssim 6\,M_{in}$, a bound actually
approached when  the interaction number $q$ grows large. More
realistically, as  $q $ is related to $t$ (or $z$) by Eq. (13) on average,
the host undergoes  $q \simeq 7$ interactions from $ z=2.5$ to $0.2$,  of which only the initial
$4$ or $5$  are effective on average; correspondingly, the masses
grow by a factor up  to  $4$. On the other hand, the
mass remains unchanged for the BHs that were never re-activated
after $z = 2.5$; so the growth of $M/M_{in}$ spans  the
range from 1 to 4 at the outmost. This may be rephrased in terms of an overall
scatter bounded by a factor 2 from the average, that is, $log \,M$ is bounded by
0.3 dex.

\section{Evolving the BHs for z $< 2.5$}

On long time scales  $t >\tau_r$ the development of the MD of the
BHs  may be viewed at as a stochastic  process that increases $M$, at
given $\sigma$  and given  halo distribution $N_h(\sigma)$; the
corresponding distribution is denoted by $N(M, \sigma, t)$.

This is  ruled by the equation
\medskip
\begin{eqnarray}
\nonumber
\partial_t N(M,\sigma,t)&=&
-\frac{\alpha}{\tau_r}N(M,\sigma,t)+\\&+&
\frac{\alpha}{\tau_r}\int\, d\mu\; p(\mu\, |\,\sigma)\,
N(M-\mu,\sigma,t)~.
\end{eqnarray}
\medskip
proposed by CV02, and used also by Yu \& Tremaine (2002), Hosokawa (2004), Menou \& Haiman (2004).
 The evolutionary rate  $\partial_t N$ is contributed by two
terms. The first describes  the BHs which interact and thereby
increase their initial mass $M$, so depleting  the number
$N(M,\sigma)\, dM$ in the mass range $(M\,-\,M+dM)$. The second
describes  the number of BHs which start from a lower mass $M-\mu$
and accrete a gas amount $\mu$, with  probability $p(\mu,\sigma)$
given by Eq. (14) or (16) for the UA or r the FCA model,
respectively. Here  $\tau_r$ is  the average time between two
subsequent interactions of a galaxy, discussed in \S 4; moreover,
$\alpha\approx 0.3$ is is the host fraction in dense environment,
corresponding to the
 $30\%$  bright galaxies residing
 in groups with membership $\geq 3 $ (Ramella et al. 1999). In
the above equation the number of BHs is  conserved, while they are
re-distributed toward larger masses; to a next approximation  number
conservation may be relaxed on adding to Eq. (18) the appropriate
source term as discussed in \S 7.

To capture the evolutionary trends  given by Eq. (18) it is
convenient to consider at first small accretion events with
$\mu/M<<1$, and  Taylor expand to second order. So, we end up with
the approximate equation
\medskip
\begin{eqnarray}
\nonumber
\partial_t N(M,\sigma,t) &\simeq& -\frac{\alpha\, \langle\mu\rangle}{\tau_r}\,\,  \partial_M
N(M,\sigma,t)\,\\&+&\,\frac{\alpha\,
\langle\mu^2\rangle}{2\tau_r}\,\,
\partial_M^2N(M,\sigma,t) .
\end{eqnarray}
\medskip
This is similar to a Fokker-Planck equation, actually one based on
the probability distribution $p(\mu\, |\,\sigma)$. The coefficient
of the first order derivative $C(M,t)\equiv
\langle\mu\rangle\,/\,\tau_r$  represents the average upward
\textit{drift} of the mass under  accretion,  while the coefficient
of the second derivative $D(M,t)\equiv
\langle\mu^2\rangle\,/\,2\tau_r$ plays the role of a
\textit{diffusion} coefficient; the averages are computed on using
the probability distribution $p(\mu \,|\, \sigma)$ given in Eq. (14)
or (16). Note that in the context of QSO evolution (Cavaliere et al.
1983, Small \& Blandford 1992),
 Eq. (19) constitutes a continuity equation for the MD
that contains also a diffusive term; correspondingly, $N(M,t)$
 not only  \textit{drifts} toward larger masses, but it is also
\textit{reshaped} reflecting the  scatter in the masses added by the
stochastic re-activations.

Here we give the analytic solution of the above equation in the
simple case when the  coefficient $D(M,t)$ is constant in time and
independent of $M$ (see Shankar 2001). We solve  for the evolution from an initial mass
distribution  $\alpha N_{in}(M,\sigma)$ of the BHs residing in
groups at $z\simeq 2.5$. We perform a transformation to Lagrangean
coordinates $M_{c} \equiv M-\int_{t_{i}}^{t}C (M,x)dx$, where the drift
term is absorbed into a total derivative; so we end up with a pure
diffusive equation. This is solved by standard methods to yield
\medskip
\begin{equation}
N(M_c,\sigma,t)=\frac{\alpha}{2\sqrt{\pi Dt}}\int\, d\xi \,
N_{in}(\xi,\sigma)\, e^{\frac{(M_{c}-\xi)^2}{4Dt}}~.
\end{equation}
\medskip
Finally, we go back  to Eulerian coordinates and obtain our solution
$N(M,\sigma,t)$. We represent in Fig. 3  the local mass distribution
$N(M,t_0)=\int d\sigma [(1-\alpha)N_{in}(M,\sigma) +
N(M,\sigma,t_0)]$, with the first term due to the dormant BHs that
do not reside in groups; both terms are integrated over the variable
$\sigma$. We use  the following parameter values: the coefficient
$C(M,t) \simeq 10^8\, (M/10^9\, M_{\odot})^{2/3}\, (t/t_0)^{-2}\;
M_{\odot}/$Gyr is derived from its definition combined with Eqs. (9)
and (12), averaged with the use of Eq. (10), and considering also
that $m\propto M_h \propto \sigma^3$ applies;  the second
coefficient $D \simeq 3\,10^{14}M_{\odot}^2/$Gyr is derived
similarly, but performing a final time average.

The solution of Eq. (19) in the form $M\, N(M,t)$ is represented by
the solid, thin line in Fig. 4. It is seen  that the evolutionary
trends described by the approximate solution are close  to those
from the full Eq. (18) where the actual mass additions $\mu/M $ are
finite; the similarity is closer in  the FCA case where the
probability in Eq. (16) is effectively confined to a narrow  range.

To solve the full Eq. (18), we use  recursive stepwise integration,
having assumed all host galaxies residing in groups (a fraction
$\alpha$ of the hosts) to undergo an interaction in the time
interval $\tau_r$. The BHs are formed with initial mass $M_{in}$ by
the end of the era $z\ge 2.5$;  each of the subsequent interactions
(labeled by the index $k$) contributes an additional mass $\mu_k$.
After $q$ interactions the BH mass is given by the sum
$M=\sum_{k=0}^{q} \mu_{k}$ of the stochastic amounts $\mu_{k}$, with
$\mu_0\equiv M_{in}$. The probability density for $\mu_k$ is given
by Eq. (14) for unconstrained accretion;  if instead the constraint
by the feedback is effective,  $\mu_k$ will have the probability
density given by Eq. (16).

Thus after $q$ interactions the solution for  $N(M,\sigma, t)$ at
the time $t$ provided by Eq. (13)  is given by
\begin{equation}
N(M,\sigma,
t)=N_h(\sigma)\big{[}(1-\alpha)P_{in}(M\,|\,\sigma)\,+\,\alpha
P_q(M\,|\,\sigma)\big{]}.
\end{equation}
Here again the first contribution is due to the dormant BHs; $P_q(M\,
|\,\sigma)$, the conditional probability to find a BH mass $M$ in a
halo with given $\sigma$, is computed with the recursive equation
\begin{equation}
P_q(M\,|\,\sigma)=\int d\mu\,\, p(\mu\, |\,\sigma)\,
P_{q-1}(M-\mu\,|\, \sigma)~.
\end{equation}
This starts out with the  conditional probability for the initial step
$k=0$, that we express  as
\begin{equation}
P_{in}(M\,|\,\sigma)=\delta [M-M_{in}(\sigma)]
\end{equation}
by continuity with the $M-\sigma$ correlation produced at the end of
the previous era $ z > 2.5$.

Each sheet in Figs. 1a and 2a represents the contribution, expressed
by Eq. (21), to the MD of relic BHs from hosts with a given velocity
dispersion $\sigma$. The contribution from hosts that never
previously interacted is represented in the form of spikes peaked around the BH
masses $M_{in}$ formed at $z\ge 2.5$. The alignment of such spikes
shows how these masses
are related to $\sigma$ after the Eqs. (4) or (6),
respectively. At $\simeq  2.5$ the regime of galaxy interactions
in groups begins; now the spikes partially drift and spread along
the mass scale, as the BHs grow by such stochastic accretion
events. The growth differs in the UA and FCA model.


In the {\itshape unconstrained}   model UA (Fig. 1a) the BHs grow by
stochastic amounts $\mu$ distributed with  the probability
$p(\mu\,|\,\sigma)$ given by Eq. (14). This   extends over the mass
axis with decreasing values.
In the {\itshape feedback-constrained} model FCA (Fig. 2a) the
growth of the BHs is reduced, being bounded by a  factor 4. This is
because the constraint cuts off the upper range of the probability
distribution $p(\mu, \sigma)$ tucking it -- as it were -- at the
upper bound of the  range of $\mu$, as described in detail by Eq.
(16).

In Figs. 1b and 2b  we project onto the $M, \; \sigma$ plane the
probability  $N(M, \sigma, z=0)/N_h(\sigma)$ given by Eq. (21); we
have also reported the data concerning $M-\sigma$ as discussed by
Tremaine et al. (2002). We predict the feedback-constrained mass
accretion to be not so abundant as to materially  change  the early
$M-\sigma$ correlation. So in our constrained model this is rooted
back at $z\gtrsim 2.5$, with only mild alterations occurring
afterwards, and these mainly at intermediate and small masses; the
result is consistent with the observations by Shields et al. (2003).

In Figs. 3 and 4 we show the behaviour  of $N(M,t)=\int
d\sigma\,N(M,\sigma,z)$, the MD integrated  over $\sigma$, i.e.,
over the sheets shown in the previous two Figures; note the
dependencies on $\sigma$ of Eq. (21). We illustrate how the MD
changes from $z=6$ (dashed lines) to $z=2.5$ (dotted lines), and
then to $z=0$ (thick solid lines); in both Figures the shaded region
represents the observational evaluations from local data in
different bands (Yu \& Tremaine 2002, Shankar et al. 2004). In the
UA model shown in Fig. 3 the relic MD at $z \simeq 0$ extends toward
large masses with a high tail; here the match to the data is poor,
since the latter indicate a sharper decline. Fig. 4 shows how in the
FCA model the feedback depletes the tail and produces a closer fit
to  the local data; the accretion so constrained goes to
contributing significantly more in the intermediate mass range where
it causes the MD to swell,  again in agreement with the data. We
have included the contribution of later accretion events occurring
in the field (to be discussed in \S 7), to the effect of  yet
improving at small masses the agreement with the observations, as
can be seen by comparison with the dashed-dotted line which does not
include this contribution. The approximate diffusive solution for
the MD is represented by the thin solid line in Fig. 4.

The same interactions that produce the MD above when time-integrated over
several Gyrs, also  produce over times  $10^{-1}$ Gyr the
LF that we discuss next.

\section{The luminous evolution for $z<2.5$}

At high redshifts $z > 3$ where  the luminosities are expected
to be Eddington limited and feedback constrained as discussed in \S 3,
the LF has been computed by CV02; we just recall in Fig. 6 their prediction
at  $z \simeq 4.5$ to show the agreement with
the recent observations by the GOODS survey (Cristiani et al. 2004b).

Here we focus on the the range $z<2.5$, an era when the supply is
\textit{limited} due to  progressive exhaustion of the  gas
reservoirs in the hosts by the many accretion episodes that produce
$L$. Two reasons concur to cause here a complex relation between LF
and MD. First, as anticipated in \S 1 an accretion episode of a mass
$\mu$ over a time $\tau$ produces the luminous output $L$ given by
Eq. (2); instead,  the cumulative mass growth is given by
$M=\sum_{k}\mu_{k}$, involving time-integration over the history of
the QSO population. Second,  the mass $\mu$ made available for the
accretion is ruled by the  interactions with their stochastic
component; so $\mu$ may be quite smaller than $M$, and the
accretions may be very sub-Eddington. In other words, the relation
of $L\propto\mu$ with $M$ is no longer tight.

The actual relation stems from the conditional probability
distribution $p(\mu\,|\,\sigma)$ for accreting $\mu$ at given
$\sigma$, expressed in Eq. (14) or (16) for the UA or the FCA model,
respectively. This yields also the conditional probability density
$p(L\,|\,\sigma)$ for the luminosity to attain in a reactivation the
value $L$ given by Eq. (2), and then fade out. So at any given time
we have
\medskip
\begin{equation}
p(L\,|\, \sigma,t)=p(\mu\,|\,\sigma)\; \frac{d\mu}{dL};
\end{equation}
\medskip
where $\mu \propto m(t)$ effectively  depends  on $t$ due to Eq.
(12).

Note that in the UA model,  the  mass $M$ accumulated at time $t$
into a BH obeys the relation $M(t) = [m_{in}-m(t)]/10$ in terms
of the residual gas mass $m(t)$ in the host; so, the distribution of
the latter is directly linked to the MD of BHs. This no longer holds
in the FCA model, where  we use  the average value for $m(t)$
given by Eq. (12).

The rate of the reactivations is  $\beta  N_{bh}/\tau_r$,  in terms
of BH number distribution $N_{bh}(\sigma,t)= \int dM\,N(M,\sigma
,t)$, and of $\beta = \alpha/ N_g \approx \alpha /3$ that represents
the fraction of hosts residing in groups (see \S 5) and interacting
over a mean time $\tau_r\simeq \, $ a few Gyrs.

The above components  can be brought together to yield the LF, upon
using the formalism of the continuity equation along the $L$ axis, as  developed by
 Cavaliere et al. (1983). This takes on the form
\medskip
\begin{equation}
\partial_tN(L,\sigma,t)+  \dot{L}\; \partial_LN(L,\sigma,t)   =
\beta \;\; \frac{N_{bh}(\sigma,t)\; p(L\,|\,\sigma,t)
}{\tau_r(t)}~,
\end{equation}
\medskip
considering for simplicity light curves equal and monotonically
decreasing on the scale $\tau$. The  solution is given by
\medskip
\begin{eqnarray}
N(L,\sigma,t) =
 \frac{1}{\dot{L}}\int_L^{\infty}
dL'\;\, \beta (t')\; \frac{ N_{bh}(\sigma,t')\; p(L'\,|\, \sigma,t')}{\tau_r(t')} ~;
\end{eqnarray}
\medskip
for the numerical computations represented in Figs. 5 and 6 we use  the
specific values $\dot{L}=-L/\tau$ with $\tau \simeq t_d = 10^{-1}$ Gyr, while $\beta$ saturates
to the value $0.1$ soon below $ z= 2.5$.

Dependencies  on $\sigma$  arise in Eq. (26) from the probability
distribution $p (\mu\,|\, \sigma)$ that enters Eq. (24), and from
the integrated $N_{bh}(\sigma,t)$ obtained from Eq. (21). Upon
convolving over $\sigma$, we obtain the bolometric LF; this is
converted to  optical luminosity $L_B$ on using the standard
bolometric correction of 10. The results are plotted  in Figs. 5 and
6 for unconstrained and feedback-constrained accretion -- our  model UA
and FCA, respectively -- and are  discussed below; we compare them with
the data of Boyle et al. (2000) and Grazian et al. (2000). As to $z >
3$, in Fig. 6 and its caption we also  recall that in our FCA model
the LF has flat shape, low normalization, and $z$ behaviour in detailed agreement
with the observations over a wide range of $L$ at $z\simeq 4 - 5$.

Points to be noted are as follows. First, the shape of the LF may be
schematically rendered  as a rather flat faint section going over to
a steeper bright section. The faint section results from the flow
along the $L$ axis (2$^{nd}$ term on l.h.s. of Eq. 25) of sources
fading from their top luminosities (stochastically provided after
Eq. 24); the bright section results from the shape of $p(L)$
convolved over $\sigma$.

Second, the evolution we expect for the LF may be understood as a
combination of a mild ``density evolution'' and a stronger
``luminosity evolution''. The former   arises as the  interactions
re-activate  the BHs in groups on a time scale $\tau_r$ that grows
moderately with cosmic time; this causes a decrease of the amplitude
of the LF by a factor around $ 2$ down to $z \simeq 0.5$. The latter
evolution occurs because on the same time scale the interactions
exhaust the gas content in the hosts according to Eq. (12); less
remaining  gas means lower luminosities following $L\propto m(z)$.

Third, the accretion history we envisage produces at late $z$ an
excess of sources compared to the optically selected AGNs; this
leaves room for sources in X-rays  and for obscured objects (Fiore
et al. 2003, Ueda et al. 2003). A larger excess is produced at
fainter $L$ by the field processes to be discussed in \S 7.

Decreasing luminosities and increasing masses produce Eddington
ratios $\lambda_E$ declining on average. In fact, the luminosities
of the sources when re-activated at  $z \simeq 0.2$ are lower than
at $z \simeq 2.5$ by a factor $(1-\langle f\rangle)^q\simeq 1/3 $
for $q=7$; meanwhile,  the masses grow at most by a factor 4 as we
have seen at the end of \S 4, so the Eddington ratios decrease
to about $1/12$ toward $z\simeq 0.2$. The actual distribution
of $\lambda_E $ at given $\sigma$ is computed on convolving the
probability for a luminosity $L$ given by Eq. (24) and that for a
mass  $M$ given by Eq. (22), with the variables combined as to yield
$\lambda_E$. This may be represented as
\begin{eqnarray}
\Pi_q(\lambda_E,\sigma)&=&\\
\nonumber \int &dM&\, \int d \mu\; P_q(M|\sigma)\; p(\mu|\sigma)\;
\delta\big{(}\lambda_E- \mu/ M \big{)} ,
\end{eqnarray}
considering that for any  value of the Eddington ratio $\mu/M =
Lt_E/ Mc^2 $ holds. Fig. 7 illustrates our  numerical results with
feedback-constrained accretion, the FCA model that enters through
the expressions of $ p$ from Eq. (16) and $P$ from Eq. (22).

The figure shows  the complex  behaviour of
 $\lambda_E $ that we find for increasing number of
interactions undergone by the hosts. In fact, we have simplified
our plots starting them from a sharp initial condition at
$z = 2.5$.
The overall decline is due to the average dimming  of all
luminosities, more rapid in hosts with smaller $\sigma$; these
undergo relatively faster gas consumption due  to the  scaling
$\langle f \rangle \propto \sigma^{-1}$ discussed in \S 4. The
overall range of  $f$ is affected by the varying lower end $f_{min}
\propto \sigma^{-1}$ and also by the distribution of $M'$; the
result is  illustrated by the pairs of lines corresponding to given
values of $\sigma$, with two instances  named in the caption. The
outcome is a correlation between $L$ and $M$ which is tighter in
smaller hosts, more sensitive (in the absence of the limits from cooling or optical depth
as discussed in \S 8) to the feedback constraint.

In sum, as $z$ decreases from $2.5$ the average values of $\langle
\lambda _E \rangle$ decline for the truly supermassive BHs in old
spheroids while the intrinsic  range covered by $\lambda _E$ widens,
consistent with the observations by McLure \& Dunlop (2003) and by
Vestergaard (2004). Next we consider  the higher values of
$\lambda_E$ contributed by interactions involving hosts located in
the field.

\section{Later accretion events in the field}

To now we have focused on  accretion events driven  for $z < 2.5$
by interactions of the old  host galaxies residing in \textit{dense}
environments like the groups. But hierarchical clustering also
envisages additional ways to feed BHs, which  for $z\lesssim 0.5$
contribute to accretion onto  BHs even when their hosts are located
in \textit{lower} density environments (the ``field").

There some new BHs continue to form by major merging events even at
$z < 0.5$, though at a reduced rate; meanwhile, starving BHs are
still refueled by interactions, though on longer time scales $\tau_f
\sim 10$ Gyr. By the same token, these later and rarer  events
involve hosts that are still moderately \textit{rich} of gas, the
latter having been depleted only or mainly by quiescent star
formation; on similar scales, satellite galaxies begin their plunge
into the hosts, importing some fresh gas.

So these later, generally smaller but still considerable accretion
events feed what is  perceived as a later population of AGNs peaking
under $z\sim 1$.  Thereby the LF is enhanced especially at faint
bolometric L, and is best observed in the X-ray band; the integrated
contribution to the BH masses is appreciable.  Since our basic
equations Eqs. (18) and (25) are linear, in our computations of the
observables we have summed the outcomes of the following three
independent processes.

i) Some major galaxy mergers still occur for $z< 2.5$, if at lower
and lower rates. These strong dynamical events may form/grow BHs
in recently reshuffled, gas-rich hosts at low $z$. Correspondingly, we add
a source term  to the r.h.s. of Eq. (18), in the form
$   P_{in}(M|\sigma)\,  \partial^+  N_{h}(\sigma,t)$ that yields
the host merging rate in terms of $ \sigma$. This
yields   some
$20\% $ of all AGNs at $z \simeq 0.5$; but these ``new" BHs  will start below the overall $M -
\sigma$ relation,  and  shine close to $\lambda_E \simeq 1$
at intermediate $L_B$.

ii) Interactions occur also in  non-virialized LSS for
$z\lesssim0.5$. Here the relative velocities $V$ are up to
$300\,$km$\,s^{-1}$, the upper bound to the pairwise ones;  the
galaxy densities $n_g$ are lower by about  $3\, 10^{-2}$, and so are
the related encounter rates $\tau_f^{-1}\simeq n_g\Sigma V$. But the
hosts so involved are about 4 times more numerous than in groups,
which include only a fraction $\alpha\simeq 0.3$ of the  galaxies in
the field (Ramella et al. 1999). The corresponding rate in LSS to be
inserted on the r.h.s. of Eq. (25) is given by $(1-\alpha)
N_{bh}(\sigma,t)\, p(L\,|\, \sigma, t)/ \tau_f(t)$;  at $z\simeq
0.5$ this contributes about  10\%   compared with  the rate in
groups, a fraction increasing  to $30 \%$  toward $z \simeq 0$ at
$L_B\gtrsim 10^{45}$erg s$^{-1}$.
On the other hand, in these conditions the gas exhaustion rate
$\dot{m}/m \simeq \langle f \rangle/ \tau_f$  is lower by $3  \, 10^{-2}$ compared with
hosts in groups, see Eq. (12); this leaves in the hosts correspondingly more residual gas, thus the
QSOs so reactivated may burst out at higher Eddington ratios than their coeval counterparts
in groups.

iii) The DM merging history also includes many events where  the
hosts end up cannibalizing  their satellite galaxies (see Menci et
al. 2003) together with the  associated, scant gaseous content; many
traces of ongoing such events are being unveiled in our Local Group,
see Martin et al. (2004); Law, Johnston \& Majewski (2005). Such
episodes cause only small accretion and weak,   often sub-Eddington
AGN emissions; these are easily drowned into the starlight or
obscured by dust in the optical band, but are noticeable in X-rays
(see Di Matteo et al. 2000). On the other hand, the capture  rate is
considerable just under $z \simeq 1 $, as the cosmic time approaches
the time scale of dynamical friction which sets the beginning of the
satellite plunge into the central galaxy.

In detail, we  consider that the satellites involved have masses
$M_s\simeq10^8\,-\,10^{10}M_{\odot}$, and begin their infall into
the host potential well under dynamical friction in standard times
$\tau_s\simeq 3\, (M_s\,/\,10^{10}\, M_{\odot})^{-0.7}$Gyr, see
Binney \& Tremaine (1987). But during the capture the gas in the
satellite is peeled off by tidal disruption and stripping (Colpi,
Mayer \& Governato 1999); so the gas masses reaching the host
nuclear region are reduced down to $\mu_s\simeq 10^{-2}M_s$. Assuming
that the initial mass function of the satellites follows the
low-mass end of the Press \& Schechter expression close to
$N_s(M_s)\propto M_s^{-2}$,  the probability distribution for
$\mu_s$ is given by
$p_s(\mu_s)=10^6\; M_{\odot}\; \mu_s^{-2}$ in the range $ 10^6 <
\mu_s < 10^8\;  M_{\odot}$.

Thus satellite captures constitute another, late  stochastic process
that contributes  to accretion.  They affect the evolution of the MD
in a way still described by an equation like  Eq. (18); but now the
average increment $\mu$
is roughly constant in time  rather than being proportional to
$m(t)$, and occurs in  many galaxies. Meanwhile, the number of
satellites is depleted  from an initial value around $10$, at the
rate
\begin{equation}
 \dot{N}(M_s)/ N(M_s) = - 1 /\tau_s ~,
\end{equation}
so that   $N(M_s)$ decreases on  scales of several Gyrs.

We illustrate in Fig.  6 how (ii) and (iii)   contribute about 70\%
to the faint end of the integral LF at $z \simeq 0.5$,  and up to
80\% at $z \simeq 0$. The latter, dominant contribution undergoes
\textit{density} evolution on the scales of several Gyrs  by the
decrease of $N(M_s)$; this will be \textit{luminosity-dependent}
owing to the basic dependence $\tau_s\propto M_s^{-0.7}$ on the
satellite mass. Meanwhile, the contribution to the integrated masses
from (ii) and (iii) is considerable, as shown in Fig. 4.

Finally, low outputs are conceivably contributed by independent
processes. In elliptical hosts the cooling of the galactic gas can
provide sufficient accretion  to power low-luminosity radio sources,
see Best et al. (2005). In spiral hosts  disc instabilities and bar
formation provide enough gas inflow to power faint optical  AGNs
(see Sellwood \& Moore 1999, Combes 2003); it will be interesting to see
what such processes contribute to the overall LF and MD.

\section{Discussion and Conclusions}

We have discussed how the standard hierarchical paradigm for
structure formation gives rise to a \textit{unified} if rich picture
for the accretion history of the supermassive BHs energizing the QSO
and AGN emissions. The paradigm envisages the host galaxies to
undergo a \textit{sequence} of dynamical events beginning with early
major mergers in high density environments, and passing over by $z
\simeq 2.5$ to milder interactions with companion galaxies in
groups;  later on,  interactions of hosts the field  join in, and
the overall sequence ends with captures from the retinues of
satellite galaxies.

All such dynamical
events trigger some gas inflow toward the nucleus, albeit
along the  sequence the accreted masses  decrease on average.
 At the two extremes of the sequence,  major mergers
and satellite captures both import fresh gas into the hosts;
while the intermediate interactions
just tap diminishing gas reservoirs by perturbing the symmetry of the host gravitational
potential and inducing non-conservation of the integrals that control
the gas equilibrium, such as  the angular momentum.

On this basis we have  computed several linked observables on using
a formalism focused on following both  the \textit{trend} and the
\textit{stochastic} components to the accretion events. We have
computed how the statistics of merging events and interactions
produce in nearly real time (i.e., within $10^{-1}$ Gyr) the
\textit{shape} of the QSO luminosity functions $N(L)$. We have also
computed how their petering out over scales of a few Gyrs  concurs
with  the exhaustion of the galactic gas reservoirs to produce
strong \textit{evolution}  in the LF described by our $N(L, z)$ in
Fig. 6. Over yet longer scales of several Gyrs their cumulative
actions add up to grow the BH masses $M$ and to \textit{change}
their distribution $N(M,z)$ as shown by Fig. 4.

 We have seen that in the early era  $z\gtrsim 2.5$,  when large proto-spheroids  were
in the process of buildup through major merging events, the BH
masses  closely tracked  those of the host haloes; correspondingly,
the MD  of the BHs also tracked closely  that of the host DM haloes,
see Eqs. (7). These conditions  hold at high $z$, when not only the
protogalactic dynamical scales are close to the Salpeter time, but
also replenishment of galactic gas is granted by the same merging
events that  trigger the  accretion; then the Eddington limit
applies to yield $L\propto M$. In turn, the MD and the LF grow
together at these high redshifts,  to yield  approximately $N(L, z)
\, dL \simeq  N(M, z)\, dM \; \tau/t$, as envisaged by Marconi et
al. (2004)  improving on Small \& Blandford (1992).

Our main focus was the later era  $z\lesssim 2.5$, when  these simple relations break down,
and $L$ takes on a different course from $M$  as does the LF from
the MD;  basically, this is  because now a supply limit applies. In fact, the hierarchical
paradigm indicates as main triggers the
interactions of the host galaxies within the small, dense groups that at these
epochs begin to virialize. These interactions, while no longer
providing fresh  gas to the host, can funnel toward the nucleus
 fractions $\mu/m$ of the residual gas  $m(z)$ left over
by previous events, that are considerable yet  lower than the BH mass already accumulated.
So in this era it is convenient to represent the
luminosities as
\begin{equation}
L \propto f \; m(z)~,
\end{equation}
to highlight   the historic  \textit{trend}
embodied in the residual gas mass $m(z)$, and  the \textit{stochastic} component  $f= \mu/m $ triggered
by the last accretion event.  At given $m$ the strength
of such  events and the levels  of the associated $L$ are
set by the stochastic distribution $p(\mu)$ related to the orbital
parameters of the interactions. The feedback constraint, when
effective, cuts off the upper  range of $\mu$ and tucks in -- as it
were -- the  distribution $p(\mu)$ at its upper end as shown by
Fig. 2a.

The result is a basic shape $N(L) \, L \propto p(\mu)\propto L^{-2}$
modified by convolution over $\sigma$, see Fig. 6. For $z < 2.5$ the
$z$-depending  LF embarks on a fast decrease; it is intrinsic to
our view that the peak of the QSO evolution should be found at
$z\simeq 2.5$ close to the beginning of the virialization era for small and dense groups, the
sites most conducive to interactions.

We stress why  physical continuity between the two main regimes of
accretion is bound to arise across $z = 2.5$, and how this is
implemented in our analytic calculations, in particular as for
$N(L,z)$ and $N(M,z)$. The first issue clearly goes back to the
smooth transition large galaxies $\rightarrow$  small groups in the
hierarchical scenario, that in particular implies $\tau_r(2.5)
\approx t_{dyn} (2.5)$. The second issue is related in the UC model
to the close equality $L \propto \mu_1 \lesssim  M_{in}$, that is
ensured when the mass actually accreted on the first interaction
satisfies $f/10 \approx M_{in}/m $; this is the case with the
average values $f \approx 0.1$ that we independently evaluate from
Eqs. (9) and (10). The related semi-analytic computations by Menci
et al. 2003  visualize such a smooth transition. On the other hand,
in our FCA model the coupling efficiency $\phi $ is clearly
continuous across z = 2.5, and this ensures the  condition $\mu_1
\lesssim M_{in}$ to hold, as shown in Sect. 4, by Eq. (17) and
preceding lines. Note that our coupling levels $\phi \simeq 10^{-2}$
(independently motivated at the end of \S 2), imply a supermassive
BH (active of dormant) to inhabit nearly all bright galaxies, with
masses satisfying the $M-\sigma$ relation. Moreover, Lapi et al.
2005 show that such values for $\phi$, when used in the feedback
balance extended  to the hot gas pervading groups,  also yields the
appropriate  $L_X - T$ relation for the associated X-ray emissions.

While the \textit{first population} of QSOs and powerful AGNs of
\textit{high} density ancestry is on its decline, at  $z \simeq 0.5$
our  LF shows an excess of faint QSOs over standard Type 1 sources;
as shown in Fig. 6, the excess relatively increases at fainter
bolometric luminosities $L \lesssim 10^{44}$erg s$^{-1}$ and toward
lower $z$, mainly due to the \textit{field} processes discussed in
\S 7. These summed excesses emerge at $z \lesssim 1 $ as a
\textit{later population} of AGNs, consistent with the observed
numbers of X-ray selected and obscured AGNs, see Hasinger (2003),
Fabian (2004) \footnote{Our referee has kindly pointed out to us
that the the new optical LFs of Richards at al. 2005, preprinted
after the submission of our MS, agree even better  with our
predicted excess  at faint  magnitudes.}

Correspondingly, albeit  in a fashion softened by its
time-integrated nature, $N(M, z)$ changes for $z<2.5$ by drifting to
larger $M$ while undergoing  considerable reshaping. As shown by
Fig. 4 the latter is in the form of swelling in the intermediate
range, related to the stochastic distribution $p(\mu)$ being  tucked
in by the feedback limit.
The overall change we
compute ends up in a shape agreeing with the local
observations;  Fig. 4 shows this to be the case throughout the
observed range, and predicts what will be observed at higher $z$.

The \textit{relic}, local mass density in BHs is computed on the
basis of Eq. (3) from our feedback-limited MD increases, and yields
a factor close to  3 from $z=2.5$ to $0$,  to attain  a local value
of $\rho_{bh} \simeq 5\, 10^5 M_{\odot}$ Mpc$^{-3}$ in agreement
with the estimates by Tremaine et al. et al. (2002) and by Shankar
et al. (2004). This includes the excess we find at intermediate $L$
and $z$ over the Type 1 AGNs;  the value goes up to $6\,10^5 \,
M_{\odot}$ Mpc$^{-3}$ on including also the late accretion processes
in the field.

Two more \textit{imprints} are left by this rich history. One  is to
be found in the \textit{steep} and \textit{tight}  relation of $M$
to  the DM velocity dispersion $\sigma$ in its upper section. In the FCA model we find
that Eq. (6) is updated to $z \approx 0$ by the  additional mass increase given by Eq. (17), to read
\medskip
\begin{equation}
M~ = ~ 1.3\, 10^8M_{\odot}\left(\frac{\sigma}{\sigma_*}\right)^5 ~,
\end{equation}
\medskip
illustrated by Fig.  2b.
This translates into $M \propto \sigma_c^{4.2 \div 4.5}$ in terms of the
velocity dispersion $\sigma_c$ of stars in the bulge (see \S 3),
which agrees with the observations  (recall that $\sigma_*=200$ km s$^{-1}$).
Here the scatter is  moderate, in fact,  bounded by 0.3 dex, first because the next recurrent events
contribute progressively less mass on average; second, because the
actual accretion  is constrained by the feedback. In fact, the
latter effect  is dominant in hosts with $\sigma \lesssim 300$ km s$^{-1}$
when the quasar output is coupled to the surrounding medium at
levels  $\phi \approx 10^{-2}$. In full, the  limit is
provided by Eqs. (15) to read $(\sigma/c)^2 < (5\, 10^{-2}\,  \eta \,
\phi\, f_{min}\,
 t_{d}\,/\tau)$; on recalling from \S 4 and \S 3 that
$f_{min}=5\,10^{-2}\, \sigma_*/\sigma$ applies,  the condition
may be recast to read
\medskip
\begin{equation}
\sigma < 300\, \left[ \frac{\eta}{10^{-1}}\,   \frac{\phi}{10^{-2}}
\, \frac {f_{min}}{0.05}\,
\frac{t_{d}}{\tau}\right]^{1/3}\left[\frac{A}{0.6}\right]\;
km\,s^{-1}.
\end{equation}
\medskip

At the other extreme of small haloes   we note that  a BH may end up
its trajectory on the $M -\sigma$ plane above the values given by
Eq. (30) if  the coupling is weaker than our standard value  $\phi
\approx  10^{-2} $; this may occur with a coupling $\phi\propto R$
proportional to an optical depth, so that the balance condition
yields at \textit{low} $\sigma$  an upturn toward $M \propto
\sigma^4 \propto \sigma_c^{3.5}$. The upturn is yet enhanced to $M
\propto \sigma^3$ with a larger scatter, when cooling faster than
the dynamical time (more likely to occur in small dense haloes)
offsets the feedback, and emulates the condition  $\phi \ll 10^{-2}$
illustrated by the Fig. 1b in its left corner,  see also CV02. The
low $\sigma$ data of Onken et al. (2004), and Barth, Greene \& Ho
(2005) apparently indicate these conditions to apply; we expect this
may  occur for $\sigma < 70$ km/s, an issue that we will develop
elsewhere.

The other imprint left by stochastic but dwindling activity in high
density environments is to be found in the Eddington ratios that we
find to be  widely \textit{scattered}, yet \textit{declining} on
average to local values $\lambda_E \lesssim 1/3$ for the most
massive BHs, see Fig. 7. The decline is due primarily to the average
decrease of the luminosities under the generally weakening
interactions that tap diminishing gas reservoirs. Intrinsic scatter
is caused primarily in $L$ and in a milder integrated form in $M$,
by the stochastic distribution of orbital parameters in the
interactions, and by the the dependence on $\sigma$ of their
effects. Under the control by the feedback the overall scatter is
actually constrained to  under  a factor $10$ with  some mixing
caused by the host dispersion $\sigma$, as shown by the lines in
Fig. 7. The result is consistent with the observations by McLure \&
Dunlop (2003) and by  Vestergaard (2004), considering  conceivable
sources of additional scatter, such as: the initial conditions for
$z > 2.5$ will itself contain some dispersion; additional scatter in
the data clearly comes at low $z$ from observational selection
picking up lower  value of $\lambda_E $; at high $z$  the current
estimates of $M$ grow more uncertain based on  scaling relations
extrapolated from nearby AGNs (see Kaspi 2000, Vestergaard 2004).

Beyond detailed modeling,  we stress that the
observed combination of  scattered but declining ratios $\lambda_E
(z)$ with the tight and steady upper $M -\sigma$ correlation indicates
supply-limited activation of BHs in stochastic but generally dwindling
accretion episodes. This is because the later and weaker
repetitions controlled  by  the feedback increase $M$ and its scatter only moderately, see Fig. 2b;
meanwhile they decrease the average $L$ and enlarge its variance considerably, to yield
\textit{widening} scatter to
$\lambda_E (z)$ superposed to an overall  \textit{declining} trend,
see Fig. 7. The two figures together illustrate how  these two trends are
made consistent by the \textit{feedback} constraint, and lead
us to favour the FCA model for  $ 70< \sigma < 300$ km s$^{-1}$ .

In sum, a \textit{sequence} of fueling modes enliven the  uniform
underlying BH paradigm. Out of the several, attendant astrophysical
processes our work has focused and linked three major components:
the dynamical events, that recur with decreasing average strength
and trigger stochastic but generally dwindling accretion episodes;
the ensuing depletion of the galactic gas reservoirs, that in dense
environments run out of the supply for accretion and for star
formation; the constraint  imposed to actual accretion by the energy
feedback from the very source emissions. Our predicted outcomes
include: flat LFs at higher $z>3$, flattening yet at fainter
luminosities; a steep and sharp  $M - \sigma$ correlation for the
truly supermassive  BHs;  an upper cutoff to their local  MD;
decline and  scatter of the Eddington ratios, widening to low $z$.
All features in telling agreement with the developing observations.

To these accretion modes that prevail in \textit{dense} environments
and power a first QSO Population, we have added (taking advantage of
the linearity of our basic Eqs. 18 and 25, see \S 7) in Figs. 6 and
7 also the independent   modes persisting or standing out for hosts
in the  \textit{field}.  There we have found a \textit{second},
later  AGN population to arise and evolve slowly simply because the
density-dependent interaction times $\tau_r = 1/n_{g} \,\Sigma\, V$
given in  Eq. (8) are considerably longer,  around 10 rather than 1
Gyr. Relatedly, high values of $\lambda_E $  still occur at $z
\simeq 0.5$ in some  $20\%$ of the faint sources and in some 70\% of
the bright ones; they arise when hosts  in the field -- still
\textit{gas-rich} just because their  encounters and interactions
are late and rare  -- get involved in strong dynamical events that
kindle up relatively small BHs starting from under the constrained
$M-\sigma$ relation (for related evidence, see Tanaka 2004).

We conclude with two straight implications of our picture. First, in
\textit{massive} spheroids  built up in dense environment, the same
rapid exhaustion of the gas reservoirs that  causes  strong
evolution in the early QSO/AGN population is also bound to cause
early \textit{reddening} of the star populations. Later and slower
field accretion modes, instead,  involve \textit{blue} galaxies
still rich of gas and actively forming  stars. All that will be
observed (cf. Kauffmann et al. 2003, Hasinger 2004) as a
\textit{bimodal} QSO-AGN population, correlated to the bimodal
colours of the galaxy population (Dekel \& Birnboim 2004). Second,
early small spheroids form small BHs with the reduced efficiency
discussed in \S 3 causing the flattened and low LF shown in Fig. ;
this means a small fraction of BHs conspicuous at early $z$. At low
$z$, instead, many blue galaxies will harbor  smaller spheroids and
smaller BHs, either limited by the feedback constraint or not yet
matured; their fueling and illumination by field triggering modes -
although basically driven by the hierarchical formation of the
underlying structures -- will be perceived as an
\textit{anti-hierarchical} development of the activity.

\section*{Acknowledgments}

We have benefited  from  fruitful discussions with A. Franceschini,
P. Rafanelli, F. Vagnetti, and especially with A. Lapi and N. Menci.
We thank our referee for  stimulating comments and suggestions,  and
for pointing out the agreement of our predicted  LF with
observations in press. Work partially supported by  INAF and MIUR.

\newpage

\begin{figure}
\includegraphics{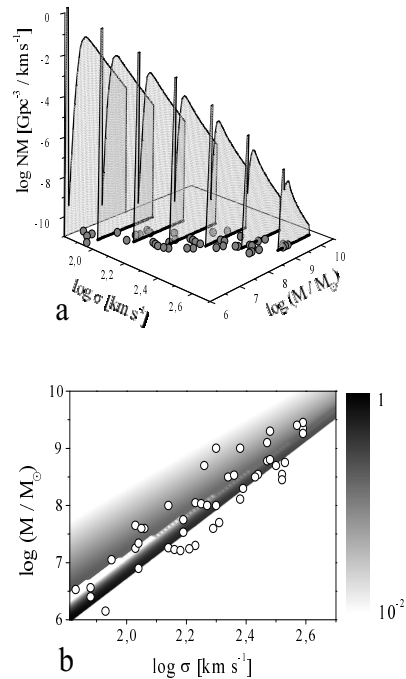} \vspace{11cm} \caption{1a. The bivariate, local
distribution $N(M, \sigma, z=0)\; M$ of the masses $M$ of BHs in
hosts with DM velocity dispersion $\sigma$, when the accretion is
unconstrained (the UA model). This is obtained after the first, more
effective 5 interactions from Eqs. (21) and (22), on specifying
$p(\mu \,|\, \sigma)$ after Eq. (14) (see the end of \S 4). The
actual distribution will be  smoothed out by the variance in the
number $q$ of interactions undergone by any given BH.} {1b.  The
local $M-\sigma$ relation, as obtained on projecting the
distribution in Fig. 1a onto the $M, \; \sigma$ plane, see \S5 of
text. The levels shown by grayscale tones are normalized to their
maximal value. Data points from Ferrarese \& Merritt (2000) and from
Gebhardt et al. (2000).}
\end{figure}

\newpage

\begin{figure}
\includegraphics{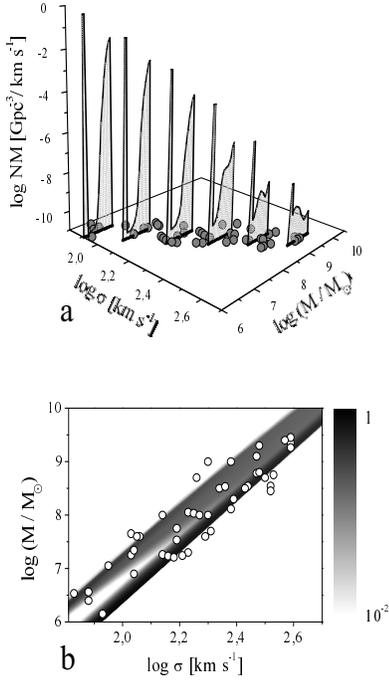} \vspace{12cm} \caption{Same as Fig. 1, but for
feedback-constrained accretion (the FCA model). Comparing with Fig.
1, note the tighter relation of the averaged $M$ with  $\sigma$. }
\end{figure}

\newpage
\begin{figure}
\includegraphics{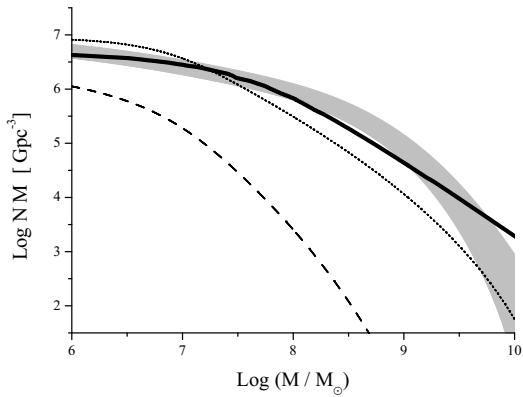} \vspace{7cm} \caption{The evolution of the BH mass
distribution  $M\, N(M, z)$ derived from Eq. (21) for the UA model, and shown  at $z=6$
(dashed line), $z=3$ (dotted line)
and $z=0$ (thick solid line).  The shaded region represents
the local estimates given by Yu \& Tremaine (2002) and by Shankar et
al. (2004).}
\end{figure}
\newpage
\begin{figure}
\includegraphics{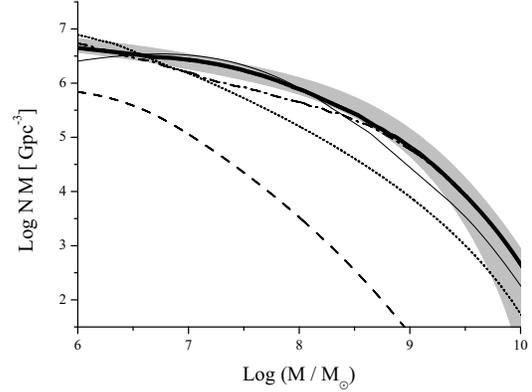} \vspace{7cm} \caption{Same as Fig. 3, but for the FCA
model. Here the thick solid line includes satellite captures, while
the dot-dashed does not. The thin solid line represents the solution
of the approximate Eq. (19); this yields a close approximation in
this case, as expected (see text,  \S 5).}
\end{figure}
\newpage
\begin{figure}
\includegraphics{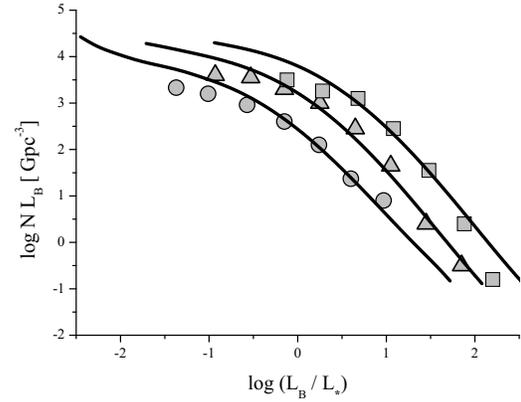} \vspace{7cm} \caption{The optical QSO luminosity
function in the form $L\, N(L,z)$  for the UA model;
from bottom to top, $z=0.5$, 1 and 2.5.
Data from Boyle et al. 2000 and Grazian et al. 2000,
marked by circles ($z=0.5$), triangles ($z = 1$), and squares ($z= 2.5$).}
\end{figure}
\newpage
\begin{figure}
\includegraphics{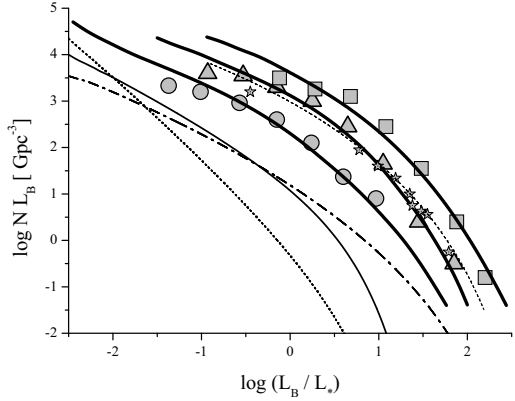} \vspace{7cm} \caption{Same as  Fig. 5, for the FCA
model. Here we have added the LFs at $z \simeq 0$:  the thinner
\textit{solid} line represents the contribution from high density
environments (the first  QSO/AGN population); the \textit{dashed}
lines represent the contributions from low density environments (the
later AGN population), with the satellites captures dominating the
faint and interactions the brighter end. Moreover, the
\textit{dotted} line represents the LF computed with the same FCA
model at $z = 4.5$, compared with the observations (represented with
stars) by  Schmidt et al. (1995); Kennefick et al. (1995); Fan et
al. (2001); Cristiani et al. (2004b).}

\end{figure}
\newpage
\begin{figure}
\includegraphics{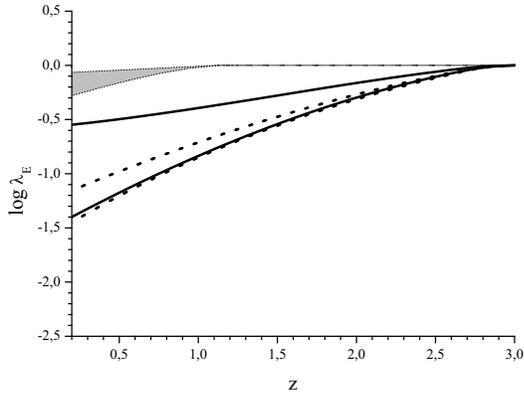} \vspace{7cm} \caption{The evolution of the Eddington
ratios $\lambda_E$ computed from Eq. (27) in the FCA model, for red
host galaxies in high-density environments, see \S 6. The ratios for
two specific values of $\sigma$ and two probability levels are
represented with different line styles: the pair of \textit{solid}
lines correspond to $\sigma =400$ km s$^{-1}$, and the
\textit{dotted} lines to $\sigma = 100$ km s$^{-1}$; each pair marks
the range from the zero of $\Pi_q$ (lower) to the $95\%$ confidence
level (upper). The redshift $z$ is  related to $ q $ as given by Eq.
(13) and by the cosmological $t-z$ relation recalled at the end of
\S 1. The upper shading represents the ratios related to blue host
galaxies in low-density environments, see \S 7. }

\end{figure}


\label{lastpage}

\end{document}